\documentclass[preprint,12pt]{JHEP3} 

\usepackage{epsfig,multicol}

\newcommand\fverb{\setbox\pippobox=\hbox\bgroup\verb}
\newcommand\fverbdo{\egroup\medskip\noindent%
			\fbox{\unhbox\pippobox}\ }
\newcommand\fverbit{\egroup\item[\fbox{\unhbox\pippobox}]}
\newbox\pippobox

\newcommand{\EQ}{\begin{equation}}
\newcommand{\EN}{\end{equation}}
\newcommand{\bea}{\begin{eqnarray}}
\newcommand{\ena}{\end{eqnarray}}
\newcommand{\bdis}{\begin{displaymath}}
\newcommand{\edis}{\end{displaymath}}

\renewcommand{\t}{\tau}

\newcommand{\pa}{\partial}

\newcommand{\nn}{\nonumber \\}

\title{ Comment on Geometric Interpretation of Ito Calculus on the Lattice }

\author{Naohito\ Nakazawa\\
     High Energy Accelerator Research Organization(KEK) \\
     Tsukuba, Ibaraki, 305-0801, Japan\\
	 E-mail: \email{naohito@post.kek.jp}}
\received{February 20, 2001} 		
\revised{May 1, 2001}
\accepted{November 27, 2001}		

\abstract{A covariant nature of the Langevin equation in It${\bar {\rm o}}$ calculus is clarified in applying stochastic quantization method to U(N) and SU(N) lattice gauge theories. The stochastic process is expressed in a manifestly general coordinate covariant form as a collective field theory on the group manifold. A geometric interpretation is given for the Langevin equation and the corresponding Fokker-Planck equation in the sense of It${\bar {\rm o}}$.} 

\begin{document} 

\newcommand{\caldslash}{{\cal D}\!\!\!\! \slash}



Wilson's lattice gauge theory ((LGT)$_d$)\cite{Wilson} is, up to now, a very unique theory which provides the constructive definition of gauge theories in the path-integral method. As expectation values of observables are concerned, the gauge fixing procedure is not necessary because of the finite volume of the gauge degrees of freedom. There exists an alternative which enables us to calculate the expectation values of observables without explicit gauge fixing procedure, so called stochastic quantization method (SQM)\cite{PW}. The application of SQM to (LGT)$_d$ was firstly discussed in relation to the lange N reduced model\cite{GL}\cite{Halpern}\cite{Sakita} as well as the numerical studies of (LGT)$_d$\cite{DDH}\cite{HH}\cite{UF}\cite{BKKLSW}. Especially, it has been shown that the integral on the group manifold is realized by the Haar measure at the equilibrium limit\cite{GL}\cite{Halpern}. In general, there are two different type of formulation for stochastic differential equations, i.e., Langevin equations, on a group manifold. One is It${\bar {\rm o}}$ type\cite{Ito} and the other, Stratonovich type\cite{Stratonovich}. The difference is essentially comes from the set up of the equal ( stochastic ) time correlation between the dynamical variables and the random noise variables. In It${\bar {\rm o}}$ calculus, the random noises are decorrelated to the equal stochastic time dynamical variables. This is a remarkable advantage to construct a collective field theory of observables from the Langevin equation of a fundamental system in the context of SQM. In this short note, in applying SQM to U(N) (LGT)$_d$, we emphasize the importance of the concept of It${\bar {\rm o}}$ calculus and its covariant nature under the general coordinate transformation on the group manifold. The geometrical meaning is given to the role of the contact term which appears in constructing a collective field theory. Especially, the Langevin equation and the corresponding Fokker-Planck equation are expressed in a manifestly general coordinate invariant form on the group manifold. 


We apply SQM to $d$-dimensional U(N) (LGT)$_d$\cite{Wilson}, 
\bea
\label{eq:U(N)action}
S [U]
= 
- \sum_{x,\ \mu > \nu } \displaystyle{\frac{\beta}{2{\rm k}N}}
{\rm Tr}( U_\mu ( x ) U_\nu ( x + \mu ) U_\mu^\dagger ( x + \nu ) U_\nu^\dagger ( x )
+ {\rm h}.{\rm c}. )   \ ,
\ena
where link variables, $U_\mu$'s,  are U(N) group elements. 
$\beta \equiv \displaystyle{\frac{N}{g^2}}$.

To write a Langevin equation for (LGT)$_d$, we intoduce the left Lie derivative ${\hat E}_A$, 
\bea
\label{eq:left-derivative1}
\displaystyle{\frac{\pa}{\pa V^A_\mu}} U_\mu
& = & 
\displaystyle{\frac{1}{{\rm k}}}{\rm Tr}\Big( 
( \displaystyle{\frac{\pa}{\pa V^A_\mu}} U_\mu ) U_\mu^\dagger t^B 
\Big) t_B U_\mu        
 = 
\displaystyle{\frac{i}{{\rm k}}}\int^1_0 ds {\rm Tr}\Big( 
{\rm e}^{isV_\mu}t_A {\rm e}^{-isV_\mu} t^B 
\Big) t_B U_\mu                           \nn  
& \equiv &
i K_A^{\ B} (x, \mu) t_B U_\mu             \ , \nn  
{\hat E}_A  (x, \mu)                    
& \equiv & 
-i L_A^{\ B} (x, \mu) \displaystyle{\frac{\pa}{\pa V^B_\mu}}      \ , \quad
K_A^{\ C} L_C^{\ B} 
 =  
L_A^{\ C} K_C^{\ B} = \delta_A^{\ B}      \ . 
\ena
We use the following convention. $t_a$'s are elements of the SU(N) algebra in the fundamental representation, $[ t_a, t_b] = if_{ab}^{\ \ c} t_c$, which satisfy ${\rm Tr}( t_a t_b ) = {\rm k}\delta_{ab}$. 
For the generators of the U(N) algebra, we define 
$[ t_A, t_B] = if_{AB}^{\ \ \ C} t_C$, 
where $t^A = t^a $ for $A=a= 1,...,N^2 -1$, 
$t^0 \equiv \sqrt{\displaystyle{\frac{{\rm k}}{N}}} {\bf 1}$. 
By definition, we have 
\bea
\label{eq:left-derivative2}
{\hat E}_A ( x, \mu ) U_\mu (x) 
& = &   
t_A U_\mu (x)     \ ,      \nn  
{\hat E}_A ( x, \mu ) U_\mu^\dagger (x) 
& = &   
 - U_\mu^\dagger (x) t_A     \ ,      \nn  
 \Big[ {\hat E}_A ( x, \mu ), {\hat E}_B ( y, \nu ) \Big] 
& = & 
-i f_{AB}^{\ \ \ C} {\hat E}_C (x, \mu) \delta_{x,y}\delta_{\mu\nu}   \ . 
\ena
%
%
The components, $L_A^{\ B}( x, \mu )$ and $K_A^{\ B}( x, \mu )$, satisfy the Maurer-Cartan equations.
\bea
\label{eq:Maurer-Cartan}
L_A^{\ C} \pa_C L_B^{\ D} - L_B^{\ C} \pa_C L_A^{\ D} 
& = & 
+ f_{AB}^{\ \ \ C}L_C^{\ D}       \ , \nn  
\pa_B K_C^{\ A} - \pa_C K_B^{\ A} 
& = & 
- f_{B'C'}^{\ \ \ A}K_B^{\ B'}K_C^{\ C'}       \ .
\ena
%
%

We define the time evolution of the link variable, 
$U_\mu ( \t+\Delta \t ) \equiv U_\mu ( \t ) + \Delta U_\mu ( \t )$ in terms of It${\bar {\rm o}}$ calculus\cite{Ito}. We use the discretized notation for its clear understanding. 
We start from the following Langevin equation\cite{DDH},  
\bea 
\label{eq:U(N)Langevin-eq1}
& {} & U_\mu (\t + \Delta \t ,\ x)   \nn
& {} & \qquad \qquad 
= 
 {\rm exp}\Big( 
 \Delta \t ( {\hat E}(\t, x, \mu )S[U( \t )] )  
 + i \Delta W_\mu (\t, x)  \Big) U_\mu (\t, x)
\ ,   \nn 
& {} & < ( \Delta W_\mu )_{ij} (\t, x ) ( \Delta W_\nu )_{kl} (\t, y )>_{\Delta W}      \nn
& {} & \qquad \qquad 
= 
2{\rm k} \Delta \t \delta_{\mu\nu} 
\delta_{il}\delta_{kj}  \delta_{xy}     \ , 
\ena
with ${\hat E} ( \t, x, \mu ) \equiv t^A {\hat E}_A ( \t, x, \mu )$. 
$\Delta W_\mu (\t, x) \equiv t_A \Delta W_\mu^A ( \t, x )$ is a noise variable defined on the site with the direction $\mu$. We notice that, from (\ref{eq:U(N)Langevin-eq1}), $U_\mu (\t, x)$ and $\Delta W_\mu (\t', x)$ are correlated only if $\t' \leq \t - \Delta\t $. In the following, the suffix \lq\lq $< {} >_{\Delta W}$ \rq\rq denotes the expectation value by means of the noise correlation in (\ref{eq:U(N)Langevin-eq1}). The equation of motion is given by 
\bea
\label{eq:U(N)equation-of-motion}
{\hat E}(x, \mu )S[U] 
= 
- \sum_{x,\ \mu > \nu } 
\displaystyle{\frac{\beta}{2N}} 
( U_\mu ( x ) U_\nu ( x + \mu ) U_\mu^\dagger ( x + \nu ) U_\nu^\dagger ( x )
- {\rm h}.{\rm c}. )      \ .
\ena
Since $ {\hat E}( x, \mu )S[U] $ is anti-hermitian, the form of the Langevin equation (\ref{eq:U(N)Langevin-eq1}) ensures that the time development keeps the link variable within the element of U(N) group. Up to the order of $\Delta \t$, we obtain the following Langevin equation,
\bea
\label{eq:U(N)Langevin-eq2}
\Big( \Delta  U_\mu (\t, x) \Big) U_\mu (\t, x)^\dagger 
 =  
 \Delta \t ( {\hat E}( \t, x, \mu )S[U( \t )] ) + i \Delta W_\mu (\t, x) 
 - \Delta\t {\rm k}N {\bf 1}  \ . 
\ena
The appearance of the term $\Delta\t {\rm k}N {\bf 1}$ is more transparent if we notice that the constraint, 
$ 
U_\mu ( \t + \Delta \t ) U_\mu ( \t + \Delta \t )^\dagger = 1
$, 
implies
\bea
\label{eq:Ito-constraint1}
\Delta U_\mu U_\mu^\dagger + U_\mu \Delta U_\mu^\dagger 
& = & 
- \Delta U_\mu \Delta U_\mu^\dagger + O ( \Delta\t^{3/2})  \nn
& = & 
- 2\Delta\t {\rm k}N {\bf 1} + O ( \Delta\t^{3/2})    \ ,
\ena
whrere ${\bf 1}$ is N$\times$N unit matrix. 

For an arbitrary function of $U_\mu$'s, $F[U]$, its time development is deduced from (\ref{eq:U(N)Langevin-eq2}), 
\bea
\label{eq:time-development-observable} 
\Delta F [U]   
& = & 
\sum_{x,\ \mu} \Big(  
 \Delta \t ( {\hat E}(x, \mu )S[U] ) 
+ i \Delta W_\mu (x, \mu) 
 \Big)^A {\hat E}_A F [U]    \nn 
& {} & 
\quad - \sum_{x,\ \mu} \Delta \t {\hat E}^A {\hat E}_A  F [U]
 - \sum_{x, \mu}\Delta\t {\rm k}N (U_\mu (x))_{ij} \displaystyle{\frac{\delta F[U] }{\delta (U_\mu (x))_{ij}}}         \ , 
\ena
where 
$({\ })^A \equiv (1/{\rm k}){\rm Tr}(({\ })t^A)$. 
In order to check the consistency of the time evolution equation, we consider the expectation value of 
$(U_\mu)_{ij}(\t,x) (U_\nu)^\dagger_{kl} (\t, y)$ 
which defines an U(N) group integral under the condition $S[U] = 0$\cite{GL}\cite{Halpern}. From (\ref{eq:time-development-observable}), the differential equation to determin the expectation value is, 
\bea
\label{eq:exercise-eq1}
& {} & 
\displaystyle{\frac{d}{d\t}} < (U_\mu)_{ij}(\t,x) 
(U_\nu)^\dagger_{kl} (\t, y)>_{\Delta W}         \nn 
& = & 
- 2{\rm k} N  < (U_\mu)_{ij}(\t,x) 
(U_\nu)^\dagger_{kl} (\t, y)>_{\Delta W} 
+ 2{\rm k}\delta_{il}\delta_{kj}\delta_{\mu\nu}\delta_{xy}  \ .
\ena
Here we have taken the limit 
$\Delta \t \rightarrow 0$. 
The solution is given by 
\bea
\label{eq:exercise-eq2}
& {} & < (U_\mu)_{ij}(\t,x) (U_\nu)^\dagger_{kl} (\t, y)>_{\Delta W}       \nn
& = & {\rm e}^{-2 {\rm k}N \t} (U_\mu)_{ij}(0,x) (U_\nu)^\dagger_{kl} (0, y)  
+ \Big( 
1 - {\rm e}^{-2{\rm k}N \t} 
\Big) \displaystyle{\frac{1}{N}} \delta_{il}\delta_{kj}\delta_{\mu\nu}\delta_{xy}      \ .
\ena
Hence we conclude that the group integral with respect to the Haar measure 
$d\mu (U)$, which is normalized by 
$
\int d\mu (U) U_{ij}U_{kl}^\dagger = 
\displaystyle{\frac{1}{N}} \delta_{il}\delta_{kj}   \ ,
$
is reproduced at the equilibrium limit\cite{GL}\cite{Halpern}. We also confirm the importance of the contribution $- \Delta\t{\rm k}N {\bf 1}$ in (\ref{eq:U(N)Langevin-eq2}) to realize the Haar measure at the equilibrium limit in the sense of It${\bar {\rm o}}$ calculus. 
While, it clearly shows that, apart from the initial value dependence, the integral measure seems to receive a stochastic time dependent scaling factor at finite stochastic time. If we assume for the integral measure at the $finite$ stochastic time to be also equal to the Haar measure, the price to pay is that the group element must pick up the factor, 
$
U ( \t + \Delta \t ) = 
 {\rm e}^{-{\rm k}N \Delta \t} U ( \t )  
$.
This fact appears in the last term on the r.h.s. of (\ref{eq:time-development-observable}), 
\bea
\label{eq:measure-eq1}
- \Delta\t {\rm k}N \sum_{x, \nu} ( U_\mu (x))_{ij} \displaystyle{\frac{\delta\ }{\delta (U_\mu (x))_{ij}}}F[U] 
= 
 \Big( F[ {\rm e}^{-{\rm k}N \Delta\t} U ] - F[U]  
\Big) \ .
\ena

In order to resolve this problem on the group integral measure at $finite$ stochastic time, we need a modification of the probability distribution.  
By assuming that the expectation value of the observable $F[U]$ is given by, 
\bea
\label{eq:probability-distribution}
< F[U] (\t) >_{\Delta W} \equiv \int d\mu (U) 
F[ U]P( \t, ( 1 - {\rm e}^{-{\rm k}N \t} )U )   \ ,
\ena
we obtain the Fokker-Planck equation, 
\bea
\label{eq:U(N)Fokker-Planck-eq1} 
\displaystyle{\frac{\pa}{\pa \t}} P( \t , U) 
= 
- \displaystyle{\frac{1}{{\rm k}}}
\sum_{x,\ \mu} {\rm Tr} \Big\{ 
{\hat E} ( x, \mu ) \Big( 
{\hat E} ( x, \mu ) 
+  ( {\hat E}(x, \mu )S[U] ) 
\Big) \Big\}  P( \t, U )          \ .
\ena
This shows that U(N) (LGT)$_d$ is reproduced at the equilibrium limit. The probability distribution $P( \t, U )$ behaves 
$
\lim_{\t \rightarrow \infty} P( \t, U ) = {\rm e}^{-S[U]}  \ 
$ by the time evolution governed by the Langevin equation (\ref{eq:U(N)Langevin-eq2}). 

Even if we chose SU(N) insted of U(N), we must require (\ref{eq:Ito-constraint1}). For SU(N), the particular contribution in the r.h.s. of (\ref{eq:Ito-constraint1}) is given by replacing $N$ to $(N^2-1)/N$. Then 
we obtain the Langevin equation for SU(N) (LGT)$_d$,   
\bea
\label{eq:SU(N)Langevin-eq1} 
& {} & ( \Delta U_\mu (\t, x ) ) U_\mu^\dagger (\t, x )  \nn 
& {} & \qquad \qquad 
= 
\Delta \t {\hat E}'( \t, x, \mu)S[U( \t )] + i \Delta {W'}_\mu ( \t, x) 
- \Delta \t {\rm k}\displaystyle{\frac{N^2-1}{N}} {\bf 1} \ ,   \nn
& {} & < ( \Delta {W'}_\mu )_{ij} (\t, x ) ( \Delta {W'}_\nu )_{kl} (\t, y )>_{\Delta W'}      \nn 
& {} & \qquad \qquad 
= 
2{\rm k} \Delta \t \delta_{\mu\nu} 
\Big( \delta_{il}\delta_{jk}  
- \displaystyle{\frac{1}{N}}\delta_{ij}\delta_{kl}
\Big) \delta_{xy}   \ ,
\ena 
where $U_\mu$ is the element of SU(N), ${\hat E}' ( x, \mu ) \equiv t^a {\hat E}'_a ( x, \mu )$ 
and 
$\Delta {W'}_\mu ( \t, x) = t_a \Delta {W'}_\mu^a ( \t, x) $. 
It deduces the following Fokker-Planck equation for the probability distribution in (\ref{eq:probability-distribution}) where $N$ in the scale factor is replaced to $(N^2-1)/N$ for SU(N) case, 
\bea
\label{eq:SU(N)Fokker-Planck-eq1}
\displaystyle{\frac{\pa}{\pa \t}} P( \t, U ) 
= - \displaystyle{\frac{1}{{\rm k}}}
\sum_{x,\ \mu}{\rm Tr} \Big\{ 
{\hat E}' ( x, \mu ) \Big( 
{\hat E}' ( x, \mu ) 
+  ( {\hat E}'(x, \mu )S[U] ) 
\Big) \Big\} P( \t, U )         \ .
\ena

Here we comment on the local gauge covariance of the Langevin equations (\ref{eq:U(N)Langevin-eq2}) and (\ref{eq:SU(N)Langevin-eq1}) . 
The local gauge transformation of the link variable $U_\mu$ is defined by 
\bea
\label{eq:U(N)gauge-transformation}
U_\mu ( \t, x )
 = 
{\rm e}^{i\Lambda ( x ) }U_\mu ( \t, x ) {\rm e}^{-i\Lambda ( x + \mu )}     \ , 
\ena
where $\Lambda (X) = \Lambda^\dagger (x)$. 
For the covariance of the Langevin equation, the noise variable defined on the site must be transformed as, 
\bea
\label{eq:SU(N)noise-transformation}
\Delta W_\mu ( \t, x )
\rightarrow {\rm e}^{i\Lambda (x)} \Delta W_\mu ( \t, x ) {\rm e}^{-i\Lambda (x)} . 
\ena
Since the noise correlations in (\ref{eq:U(N)Langevin-eq2}) and (\ref{eq:SU(N)Langevin-eq1}) are invariant under the transformation (\ref{eq:SU(N)noise-transformation}), the Langevin equations and the noise correlations manifestly preserve the local gauge symmetry. We notice that the Fokker-Planck equations, (\ref{eq:U(N)Fokker-Planck-eq1}) and (\ref{eq:SU(N)Fokker-Planck-eq1}), are also invariant under the local gauge transformation. These results are essentially not new. 
In the previous works\cite{DDH}\cite{GL}\cite{Halpern}\cite{Sakita}, the contribution such as the third term in the r.h.s. of (\ref{eq:U(N)Langevin-eq2}) has not been discussed seriously, however, it is a manifestation of It${\bar {\rm o}}$ calculus which comes from the constraint (\ref{eq:Ito-constraint1}). One might suspect that the time development described by the Langevin equation (\ref{eq:U(N)Langevin-eq2}) would leave the link variable away from the element of U(N), however, the term 
$\Delta \t {\rm k}N {\bf 1} $ 
is actually the necessary contribution which keeps the link variable within the  U(N) group up to the order of $(\Delta \t )^{3/2}$. In the following, we find that, in fact, the concept of It${\bar {\rm o}}$ calculus shows up the internal geometry in (LGT)$_d$.


The geometrically covariant nature of the Langevin equation in the sense of It${\bar {\rm o}}$ calculus was first pointed by Graham\cite{Graham} in the analysis of a stochastic process, Brownian motion of a particle, on curved spaces. Here, to introduce a geometric interpretation for the Langevin equation of (LGT)$_d$, we define a metric on the group manifold, 
$ G^{AB} ( x, \mu ) \equiv L_C^{\ A} L_C^{\ B}$ and 
$ G_{AB} ( x, \mu ) \equiv K_A^{\ C} K_B^{\ C}$. 
$G$ denotes ${\rm det}G_{AB}$. The metric is defined on each link, namely a bi-local quantity. 
 By using the components of the Maurer-Cartan one-form, the local gauge transformation (\ref{eq:U(N)gauge-transformation}) is expressed as, 
\bea
\label{eq:U(N)gauge-transformation2}
\delta V^A_\mu ( \t, x )
= \Lambda^B ( x ) L_B^{\ A} ( \t, x, \mu )- L^A_{\ B} ( \t, x, \mu ) \Lambda^B ( x + \mu ) \ , 
\ena
where $L^A_{\ B} = ( L_B^{\ A} )^t$. The continuum limit is taken by introducing the lattice spacing $\epsilon$, 
$V^A_\mu \rightarrow \epsilon V^A_\mu$. The transformation (\ref{eq:U(N)gauge-transformation2}) corresponds to the covariant derivative of $\Lambda$ at the continuum limit. For the metric $G^{AB}$ and $G_{AB}$, the variation $\delta V^A_\mu$ in (\ref{eq:U(N)gauge-transformation2}) satisfies the Killing vector equation in the \lq\lq superspace \rq\rq $\{ V^A_\mu,\ G_{AB}\}$, 
\bea
\label{eq:Killing-vector-eq1}
\delta G^{AB} 
& \equiv & 
\displaystyle{\frac{\pa G^{AB}}{\pa V^C_\mu}} \delta V^C_\mu  \ , \nn
& = & 
G^{AC} \displaystyle{\frac{\pa \delta V^B_\mu}{\pa V^C_\mu}}  +  
G^{CB} \displaystyle{\frac{\pa \delta V^A_\mu}{\pa V^C_\mu}}  \ . 
\ena
The relation is easily derived by the Maurer-Cartan equation (\ref{eq:Maurer-Cartan}). 
By the Killing vector relation, one can lift up the symmetry property in $d$ dimensional system to $d+1$ dimensions in SQM. It is also possible to introduce the BRS symmetry to the $d+1$ dimensional system\cite{Nakazawa}. Since it is not the purpose of this short note and the extension is straightforward, we do not mention more on this issue. 

In order to describe the internal geometry on the lattice, we derive the Langevin equation for the gauge field $V_\mu (x)$. We assume that it includes the lattice spacing, the UV cut-off parameter, $\epsilon$. The r.h.s. of (\ref{eq:U(N)Langevin-eq2}) is evaluated, 
\bea
\label{eq:U(N)contact-term-eq1}
( \Delta U_\mu ) U_\mu^\dagger 
=  
\Delta V^A_\mu (\pa_A U_\mu ) U_\mu^\dagger + {1\over 2}\Delta V^A_\mu \Delta V^B_\mu ( \pa_A \pa_B U_\mu ) U_\mu^\dagger  + O(\Delta\t^{3/2})          \ . 
\ena
The equation is inverted to deduce the Langevin equation for $V^A_\mu ( \t, x ) $. 
\bea
\label{eq:U(N)Langevin-eq3} 
& {} & \Delta V^A_\mu ( \t, x )   \nn
& =  & 
- i ( ( \Delta U_\mu ) U_\mu^\dagger )^B L_B^{\ A}
+  {i\over 2}\Delta V^B_\mu \Delta V^C_\mu (( \pa_B \pa_C U_\mu ) U_\mu^\dagger )^{A'}  L_{A'}^{\ A} + O(\Delta\t^{3/2})        \nn
& = &
 \Big( - i \Delta \t {\hat E}( \t, x, \mu )S[U] +  \Delta W_\mu ( \t, x )   
 + i  \Delta\t {\rm k}N {\bf 1} 
 \Big) ^B L_B^{\ A}         \nn
& {} & \qquad + 
i \Delta \t L_{B'}^{\ B} L_{B'}^{\ C} \Big( ( \pa_B \pa_C U_\mu ) U_\mu^\dagger \Big) ^{C'} L_{C'}^{\ A}   + O(\Delta\t^{3/2})       \ ,  \nn 
& =  & 
 \Big(  - i \Delta \t {\hat E}( \t, x, \mu )S[U] +  \Delta W_\mu ( \t, x )   \Big) ^B L_B^{\ A} ( \t, x, \mu )         \nn
& {} & \qquad +  \Delta \t ( \pa_B L_C^{\ A} ) L_C^{\ B} ( \t, x, \mu ) 
  + O(\Delta\t^{3/2})     \ .
\ena
The contribution 
$\Delta \t ( \pa_B L_C^{\ A} ) L_C^{\ B}$ 
in (\ref{eq:U(N)Langevin-eq3}) plays an essential role for the covariance of the Langevin equation. 

The key observation to understand a precise covariant nature of the Langevin equation for U(N) (LGT)$_d$ is that $\Delta V^A_\mu$ is $not$ a covariant quantity in It${\bar {\rm o}}$ calculus. 
Under the general coordinate transformation on the group manifold, 
$
V^A_\mu \rightarrow {V'}^A_\mu 
$, 
(\ref{eq:U(N)Langevin-eq3}) shows that $\Delta V^A_\mu (x)$ is transformed  as 
\bea
\label{eq:general-coordinate-tr}
\Delta V^A_\mu \rightarrow ( \pa_B {V'}^A ) \Delta V^B_\mu 
+ \Delta \t G^{BC}\pa_B \pa_C {V'}^A_\mu  \ .
\ena
This is not a disadvantage of It${\bar {\rm o}}$ calculus. 
From the metric tensor, we define a covariant form of $\Delta V^A_\mu$, 
\bea
\label{eq:covariant-derivative}
\Delta_{\rm cov} V^A_\mu \equiv \Delta V^A_\mu 
+ \Delta \t \Gamma^A_{\ \ BC}G^{BC} \ ,
\ena
which is a contra-variant vector under the general coordinate transformation. 
The second term yields
$
\Delta \t \Gamma^A_{\ \ BC}G^{BC}  
=
- \Delta \t 
\displaystyle{\frac{1}{\sqrt{G}}} \pa_B ( \sqrt{G} G^{AB} ) 
= 
- \Delta \t ( \pa_B L_C^{\ A} ) L_C^{\ B}     \ .
$
 Now the covariance is transparent if we write the Langevin equation (\ref{eq:U(N)Langevin-eq3}) and the noise correlation as follows.
\bea
\label{eq:U(N)Langevin-eq4}
\Delta V^A_\mu ( \t, x )
& = & 
-\Delta \t G^{AB} ( \t, x, \mu )\displaystyle{\frac{\pa S}{\pa V^B_\mu (x)}} ( \t )    \nn
& {} & \quad - 
\Delta \t \Gamma^A_{\ \ BC}G^{BC} ( \t, x, \mu )
+ \Delta_W V^A_\mu ( \t, x )    \ , \nn
< \Delta_W V^A_\mu ( \t, x )\Delta_W V^B_\nu ( \t, \nu )>_{\Delta W} 
& = & 
2 \Delta \t G^{AB}( \t, x, \mu ) \delta_{\mu\nu}  \delta_{xy}  , 
\ena
where we have introduced a collective noise field 
$ \Delta_W V^A_\mu ( \t, x ) = \Delta W^B_\mu ( \t, x ) L_B^{\ A} ( \t, x )$. The corresponding Fokker-Planck equation also appears in a manifestly general coordinate invariant form.
\bea 
\label{eq:U(N)Fokker-Planck-eq2}
\displaystyle{\frac{\pa}{\pa \t}} P( \t, V ) 
= \sum_{x,\ \mu}
\displaystyle{\frac{1}{\sqrt{G}}} \displaystyle{\frac{\pa\ }{\pa V^A_\mu (x)}} 
\Big( 
\sqrt{G}G^{AB} ( x, \mu )( \displaystyle{\frac{\pa\ }{\pa V^B_\mu (x)}} + 
 \displaystyle{\frac{\pa S}{\pa V^B_\mu (x)}} )
 P( \t, V )  \Big)         \ ,
\ena
where $P(\t, V)$ is a scalar probability density defined by,  
\bea
\label{eq:probability-distribution2}
<F[V](\t)>_{\Delta W} = \int\!\! F[V]P(\t, V)\sqrt{G}dV \ . 
\ena
The time development of the U(N) group element is reproduced by the definition 
$
U_\mu ( \t, x ) = {\rm e}^{i V_\mu ( \t, x )} 
$. 
The Fokker-Planck equation (\ref{eq:U(N)Fokker-Planck-eq2}) is derived under the assumption (\ref{eq:probability-distribution2}) in which the integral is defined by the Haar measure 
$
d\mu (U) = \sqrt{G}dV
$ 
 and the scalar probability density does not include the stochastic time dependent scaling factor. 

%
%
%


In conclusion, 
our main claim is the Langevin equations (\ref{eq:U(N)Langevin-eq2}) and (\ref{eq:U(N)Langevin-eq4}) for U(N) (LGT)$_d$. For SU(N) case, we have obtained similar results. We again emphasize the importance of the contribution $\Delta\t {\rm k} N{\bf 1}$ in the Langevin equation (\ref{eq:U(N)Langevin-eq2}) which is a particular consequence of It${\bar {\rm o}}$ calculus. Without it, we would have arrived at neither the group integral defined by the Haar measure (\ref{eq:exercise-eq2}) nor the manifestly general coordinate covariant form of the Langevin equation (\ref{eq:U(N)Langevin-eq4}). We have also shown that the Fokker-Planck equation are expressed in general coordinate invariant form (\ref{eq:U(N)Fokker-Planck-eq2}). It can be derived from (\ref{eq:U(N)Fokker-Planck-eq1}) by using the definition of the left Lie derivative. 
We, however, notice that the covariant nature of the Langevin equation is a direct consequence of It${\bar {\rm o}}$ calculus. While the invariant Fokker-Planck equation is derived from the assumption, the existence of the appropriate probability distribution, defined in (\ref{eq:probability-distribution}). This assumption is inevitable because we would like to define a probability distribution on the group integral with Haar measure which is not realized in a precise sense at finite stochastic time.  

The Langevin equation (\ref{eq:U(N)Langevin-eq2}) can be regarded as a collective field theory if we consider the Langevin equation (\ref{eq:U(N)Langevin-eq4}) as a fundamental one. Conversely, 
(\ref{eq:U(N)Langevin-eq4}) can be regarded as a collective field theory constructed from (\ref{eq:U(N)Langevin-eq2}). It is an attractive observation that the manifest geometric structure is realized as a collective field theory of an underlying system. The structure which is observed for the Langevin equation of (LGT)$_d$ in (\ref{eq:U(N)Langevin-eq4}) also appears in stochastic quantization of $N=1$ super Yang-Mills theory in superfield formalism\cite{Nakazawa2}.

\bigskip

\noindent
{\bf Acknowledgements}

The author would like to thank all members in theory group at KEK for hospitality. 
The work is supported in the early stage by the Ministry of Education, Science and Culture of Japan, Grant-in-Aid for Scientific Research (B), No.13135216.



\begin{thebibliography}{999}
%
\bibitem{Wilson} K. G.Wilson, \prd{10}{1974}{2445}.
%
\bibitem{PW} G. Parisi and Y. Wu, Sci. Sin. {\bf 24}(1981) 483.
%
\bibitem{DDH} I. T. Drummond, S, Duane and R. R. Horgan, \npb{220}{1983}{119}.
%
\bibitem{GL} A. Guha and S. C. Lee, \prd{27}{1983}{2412}; 
             \plb{134}{1984}{216}.
%
\bibitem{Halpern} M. B. Halpern, \npb{228}{1983}{173}
%
\bibitem{Sakita} B. Sakita, Proceedings of the 7th Johns Hopkins Workshop, ed. G. Domokos and S. Kovesi-Domkos, ( World Scientific, Singapore, 1983) 115.
%
\bibitem{HH} H. W. Hamber and U. M. Heller, \prd{29}{1984}{928}.
%
\bibitem{UF} A. Ukawa and M. Fukugita, \prl{55}{1985}{1854}. 
%
\bibitem{BKKLSW} G. G. Batrouni, G. R. Katz, A. S. Kronfeld, G. P. Lepage, B. Svetitsky and K. G. Wilson, \prd{32}{1985}{2736}.
%
\bibitem{Ito} K. Ito, Proc. Imp. Acad. {\bf 20}(1944) 519;\nn
         For example, see \lq\lq Srochastic Differential Equations and Diffusion Processes \rq\rq 2nd ed., ed. N. Ikeda and S. Watanabe ( North-Holland/Kodansha,1989). 
%
\bibitem{Stratonovich} R. L. Stratonovich, Conditional Markov process and their application to the theory of optimal control ( Elsevier, NewYork, 1968).
%
%
\bibitem{Graham} R. Graham, \pla{109}{1985}{209}. 
%
%
\bibitem{Nakazawa} N. Nakazawa, \npb{335}{1990}{546}.
%
\bibitem{Nakazawa2} N. Nakazawa, hep-th/0302138. 
%
%
\end{thebibliography}
\end{document}